\documentclass[aps,pra,twocolumn,groupedaddress]{revtex4-1}

\usepackage{graphicx}
\usepackage{dcolumn}
\usepackage{bm}
\usepackage{amsmath}
\usepackage{amssymb}
\usepackage{latexsym}
\usepackage{epsfig}
\usepackage{amsbsy}
\usepackage{array}
\usepackage{amssymb}
\usepackage{setspace}
\usepackage{bm}

\def\sint{\ifmmode{- \!\!\!\!\!\! \int}
    \else{\hbox{$- \!\!\!\! \int \ $}}\fi}

\begin{document}


\title{Analysis of interference in attosecond transient absorption in adiabatic condition}

\author{Wenpu Dong}
\author{Yongqiang Li}
\author{Xiaowei Wang}
\author{Jianmin Yuan}
\author{Zengxiu Zhao}
\thanks{Author to whom correspondence should be addressed; electronic mail: zhao.zengxiu@gmail.com}
\affiliation{Department of Physics, National University of Defense Technology, Changsha, Hunan, 410073, China}
\date{\today}

\begin{abstract}

We simulate the transient absorption of attosecond pulses of infrared-laser-dressed atoms by considering a three-level system with the adiabatic approximation. The delay-dependent interference features  are investigated from the perspective of the coherent interaction processes between the attosecond pulse and the quasi-harmonics. We  find that many features of the interference fringes in the absorption spectra of the attosecond pulse can be attributed to the coherence phase difference. However, the modulation signals of laser-induced sidebands of the dark state is found related to the population dynamics of the dark state by the dressing field.

\begin{description}
\item[PACS numbers]
32.80.Qk, 32.80.Rm, 42.50.Hz, 42.50.Md
\end{description}
\end{abstract}



\maketitle

\section{Introduction}\label{S1}  
Attosecond pulses, the shortest bursts of light ever produced~\cite{atto1,atto2,atto3}, allow for probing the dynamics of bound electrons on their natural time scales, such as attosecond streaking~\cite{streaking1,streaking2} and electron interferometry~\cite{interferometry1,interferometry2}. With the generation of the single high-quality attosecond pulse, it has been utilized as a tool for investigating the dynamical (transient) light-matter interactions on the few-femtosecond and even sub-laser-cycle timescale~\cite{45,16,12,14,11,5,7,8,9,10,13,15,17,18,19,20,21,23,24,6,40}. Recently, attosecond transient absorption (ATA) is applied to probe the valence electrons in atomic krypton ions generated by a controlled few-cycle laser field~\cite{45}. In the follow-up experiments, the pump-probe techniques, a combination of attosecond pulse and few-cycle laser field, are performed in different atomic systems, such as helium~\cite{6,40}, neon~\cite{16}, argon~\cite{12} and krypton~\cite{45}, raising many interesting features in the absorption spectra. For example, one observes periodically shifted and broadened lines of low-lying states, the interference fringes phenomena, and the Autler-Townes splitting of helium~\cite{1,6,40}, which is pumped by an attosecond extreme ultraviolet pulse (XUV) and then probed by a delayed infrared laser. Here, the information of complicated interactions between the two fields and the system is recorded in the ATA spectra.

In parallel to these experimental developments, there are several theoretical investigations how to gain fundamental insight into ATA. For the absorption of an isolated attosecond pulse in the vicinity of the helium, perturbation theory has been used to investigate the subcycle shifts and the broadened lines of $1s3p$ and $1s4p$~\cite{14}, and to yield a qualitative guide of the line shapes evolution of resonant absorption lines and energy shifts of $1s^{2}$-$1snp$~\cite{41}. Even though the perturbation theory can demonstrate the instantaneous responses of the bound electrons to the perturbing laser field, it cannot describe well the interference fringes features, the Autler-Townes splitting, and some other phenomena in ATA spectra arising from nonlinear processes. ATA spectra is also studied via a few-level model system~\cite{350,10} in the rotating-wave approximation (RWA). Although RWA contains many effects of interaction between the fields and the level system, the partial features of the delay-depended interference fringes were not reproduced within this approximation. The interference fringes in ATA spectra of helium~\cite{6}, result from the coherent quantum paths that lead to the same dipole excitation: one direct pathway is excited by the attosecond pulse and another indirect pathway is multi-photon transition driven by IR field. In Ref.\cite{11}, time-dependent Schr\"{o}dinger equation (TDSE) is used to analysed the effects of the coherent pathways on the interference fringes, and one obtains the phases of the interference fringes are dependent on the delay and the multi-photon transition driven by IR-field. However, one encounters difficulties for analyzing data from the TDSE calculation in SAE approximation, and it is still unclear for the phase offset due to the multi-photon transition.

In this paper, we focus on the delay-dependent features of the interference fringes in ATA spectra, for the sake of obtaining a simple picture of the ATA processes modulated by the IR dressing field.
In theoretical method, we explore the dependence of the laser-dressed dipole response on delay, based on a three-level system which effectively models the delay-dependent interference features contained in its ATA spectra. We also take the adiabatic approximation into account, which is widely used in the stimulated Raman adiabatic passage~\cite{26,27,28,29,30,31,32,37,38,34}, and here neglects the resonance-transition processes induced by the IR dressing field allowing us to obtain an analytical solution of the dipole response. By simulating the laser-dressed dipole response of the three-level system, we find that the quasi-harmonics~\cite{23,24} coherently interact with the attosecond XUV pulse at the energy. The phase differences between the quasi-harmonics and the attosecond XUV pulse as a function of delay, determine the periodic absorption at "$1s2p\pm 2\omega_{IR}$" in the ATA spectra.

The paper is organized as follows: In Sec.~\ref{S2}, a brief overview of theoretical methods about the three-level system and the calculation of the absorption spectra are given. We present the detail about describing the interference fringes by the coherence phase difference in Sec.~\ref{S3}, and the modulation signal of the sideband of the dark state in Sec.~\ref{S32}. Then a short summary is given in Sec.~\ref{S4}.

\section{method}\label{S2}  

\begin{figure}
\centering
\begin{minipage}[h]{0.5\textwidth}
\centering
\includegraphics[width=2in]{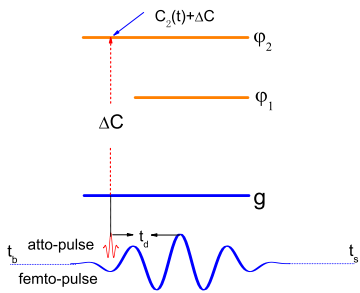}
\end{minipage}
\caption{Sketch of the process of the attosecond pulse transition to the three-level system. At the delay time $t_{d}$, the attosecond pulse populates a population $\Delta C$ from the ground state to state $|\varphi_{2}\rangle$, bringing the probability amplitude $C_{2}$ of $|\varphi_{2}\rangle$ with a instant-changed $\Delta C$. \label{f1}}
\end{figure}

To mimic near resonant $1s^{2}$-$1s2p$ XUV absorption processes in the IR-laser-dressed helium atom, we consider a three-level system interacting with a time-delayed attosecond XUV pulse and a few-cycle IR laser field. It involves the field-free states (see Fig.~\ref{f1}): the ground state $|g\rangle$, and the two excited states $|\varphi_{1}\rangle$ and $|\varphi_{2}\rangle$. $|\varphi_{2}\rangle$ is referred as $1s2p$ of helium. And $|\varphi_{1}\rangle$, whose the dipole transition to $|g\rangle$ is prohibited, is the so-called dark state. Taking the ground state energy as reference, the two excited states have energies of $\omega_1$ and $\omega_2$ respectively.

We consider the situation that the weak attosecond XUV pulse has a central frequency resonant with the excitation energy $\omega_2$, which is much larger than the frequency $\omega_L$ of the IR-dressing laser. Therefore, at each delay time $t_d$, the weak attosecond XUV pulse serves as a pump pulse to induce a resonant transition from the ground state to the excited state $|\varphi_{2}\rangle$~\cite{10,41}.

In addition we assume that $|\varphi_{1}\rangle$ to $|\varphi_{2}\rangle$ are strongly dipole coupled with the transition matrix element $d_{12}$ much larger than $d_{g2}$ that from the ground state to $|\varphi_{2}\rangle$. Then it allows us to take only the two excited states into account after the initial XUV attosecond pulse excitation, during the propagation processes of the excited three-level system in the presence of the IR-dressing laser. Expanding the wave packet excited by the attosecond pulse in terms of the field-free states $\Phi=e^{-i\omega_{1}t}C_{1}|\varphi_{1}\rangle+e^{-i\omega_{1}t}C_{2}|\varphi_{2}\rangle$, the time-dependent Schr\"{o}dinger equation for the two levels can be formulated as

\begin{figure*}
\centering
\setlength{\abovecaptionskip}{0pt}
\setlength{\belowcaptionskip}{0cm}
\begin{minipage}[h]{1\textwidth}
\centering
\includegraphics[width=7in]{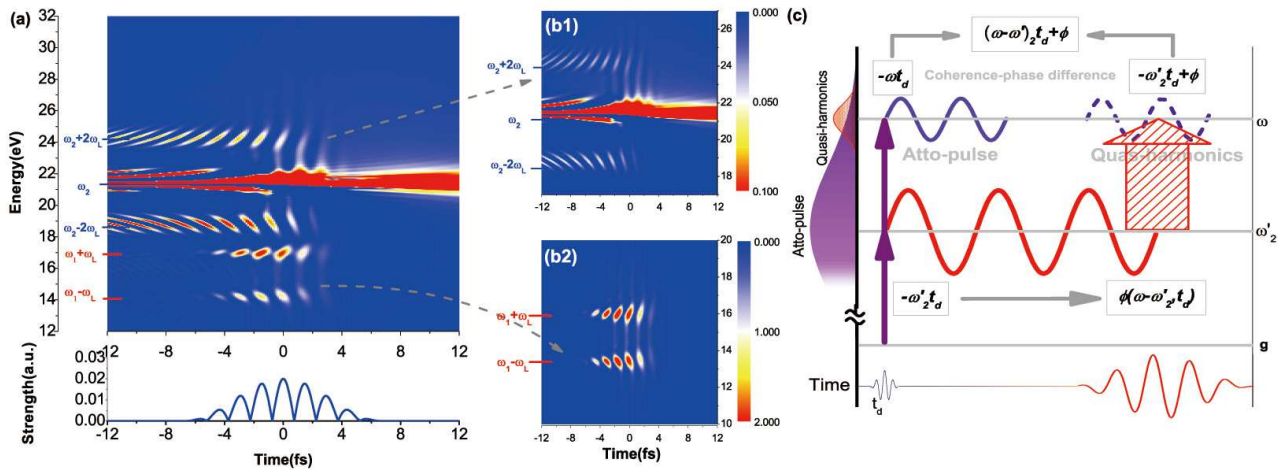}
\end{minipage}
\caption{(color online)\textbf{(a,b)} The response function $S(\omega,t_{d})$ of the three-level system as a function of the delay time. \textbf{(a)}Spectra from the numerical calculation of the three-level TDSE with the attosecond pulse ($240as$, centered in at $21.4eV$) in the presence of the dressing field (blue line, in absolute value). \textbf{(b)}Spectra from the calculation of the adiabatic model, shown in Eq.~(\ref{e3}), \textbf{(b1)} and \textbf{(b2)} are the parts relate to the quasi-harmonics from the laser-dressed dipole responses of $|\varphi_{2}\rangle$ and $|\varphi_{1}\rangle$ to the ground state, respectively. \textbf{(c)} Sketch of the coherence-phase difference. The interference fringes seen in \textbf{(a)} and \textbf{(b)} at label $\omega_{2}\pm2\omega_{L}$ result from the coherent interaction between the attosecond pulse and the quasi-harmonics. The coherent phase of the quasi-harmonics associate with the effects of the attosecond pulse and the IR-dressing field. The system obtains a initial phase $-\omega'_{2}t_{d}$ excited by the attosecond pulse at delay time, and the quasi-harmonics will be added a phase offset in the dressed processes by the IR-dressing field.
\label{f2}}
\end{figure*}

\begin{eqnarray}
i\frac{d}{dt}\left(
                     \begin{array}{c}
                       C_{1} \\
                       C_{2} \\
                     \end{array}
                   \right)=\left(
                     \begin{array}{cc}
                       0 & \gamma \\
                       \gamma^{*} & \Delta\omega_{12}\\
                     \end{array}
                   \right)
\left(
  \begin{array}{c}
    C_{1} \\
    C_{2} \\
  \end{array}
\right),
\label{eq:e1}
\end{eqnarray}
where $\gamma(t)=E(t)d_{12}$, $E(t)$ is IR-dressing field given by $E=E_{0}\varepsilon(t)\cos(\omega_{L} t)$ with the dressing field envelop $\varepsilon(t)=\cos^{2}(\frac{t\pi}{\tau})$ for $t_{on}\leq t\leq t_{off}$, otherwise $\varepsilon(t)=0$ with $t_{on}=-\frac{\tau}{2}$, $t_{off}=\frac{\tau}{2}$, and $\tau$ is the pulse duration. $\Delta\omega_{12}=\omega_2-\omega_1$ is the resonance-transition energy between the two excite states. Atomic units are used throughout unless indicated otherwise. By diagonalizing the hamiltonian at each instant, we have the dressed states $|\varphi_{+}\rangle$ and $|\varphi_{-}\rangle$, with the probability amplitudes of the dressed state are given by:
\begin{equation}\label{e2}
C_{\pm}=b_{\pm}C_{1}+a_{\pm}C_{2},
\end{equation}
where the transform coefficients are given by $a_{\pm}=\omega_{\pm}/\sqrt{\gamma^{2}+\omega^{2}_{\pm}}$, $b_{\pm}=\gamma/\sqrt{\gamma^{2}+\omega^{2}_{\pm}}$, with $\omega_{\pm}=\frac{1}{2}(\Delta \omega_{12} \pm \sqrt{\Delta \omega^{2}_{12}+4\gamma^{2}})$, the energies of the dressed states $|\varphi_{\pm}\rangle$. The dressed energies $\omega_{\pm}$ are depended on instantaneous strengths of the dressing field. In this adiabatic model, the wave packet would be strongly dressed after $t_{d}$ in the presence of the IR-dressing field and keep stable populations in the dressed states. The adiabatic model is tenable with the adiabatic condition: $|\Delta\omega_{12}|\gg\omega_{L}$ and the intensity of the dressing field is weak($\leq 10^{14}$ W/cm$^{2}$). Assuming at $t_d$, the attosecond pulse induces an instant change of the probability amplitude $\Delta C$ of $|\varphi_{2}\rangle$, then the probability amplitudes of the two dressed states are given by $C_{\pm}(t_{d})=\Delta Ca_{\pm}(t_{d})$. In the subsequent evolution, we have
\begin{eqnarray}
         C_{2}(t)=C_{-}(t_{d})a_{-}(t)e^{-i\int^{t}_{t_{d}}\omega_{-}dt}+C_{+}(t_{d})a_{+}(t)e^{-i\int^{t}_{t_{d}}\omega_{+}dt},
\label{e3}
\end{eqnarray}
the details are given in Appendix~\ref{Ap1}. Based on the time evolution of the system, in particular, we focus on dipole response from $|\varphi_{2}\rangle$ to the ground state with respect to the pump-probe delay. The dipole response in time domain is given by: $d(t)\approx 2Re[e^{-i(t-t_{d})\cdot\omega_{1}}C_{2}(t)d_{g2}]$, where we have assumed the probability amplitude of the ground state changes little, and the $C_{2}(t)=0$ before $t_{d}$ and after then $C_{2}(t)$ varies as described by Eq.(\ref{e3}).

From the induced dipole moment, the absorption of the attosecond pulse by the system can be described by a frequency-dependent response function, following the approach in Ref.~\cite{11,41,44}:
\begin{equation}\label{e4}
S(\omega,t_{d})=2Im[E^{*}_{atto}(\omega)d(\omega)],
\end{equation}
where $E_{atto}(\omega)$ and $d_{\omega}$ are the Fourier transforms of the attosecond pulse $E_{atto}(t)$ and the dipole moment $d(t)$ with delay $t_{d}$, respectively. We use the Hanning window to eliminate the noise signal resulting from the break of the dipole moment at the ends. The responds function $S(\omega,t_{d})$ tells that when the spectra phase $\varepsilon(\omega)$ of the attosecond pulse have a $\frac{\pi}{2}$ phase difference with the dipole response phase $d(\omega)$, the fields would apply a positive work on the electron and the energy would transition from the pump-probe fields to the level system. A revers process also happens with the phase difference added by $\pi$. It is a coherent interaction between the pump-probe fields and the dipole response of the level system, depending on the phase difference.


\section{Results and discussions}\label{S3}  

In this section, we calculate the ATA spectra by the adiabatic model, and compare it with the results from the numerical solution of three-level TDSE. After verifying the reliability of the adiabatic model against TDSE, we will utilized it to analyze the coherent interactions between the attosecond pulse and the delay-dependent dipole response in more details.

The parameters used in the three-level model are given as follow. The ground state energy is taken as zero, and the field-free energy $\omega_{2}$ of $|\varphi_{2}\rangle$ is $21.4$ eV corresponding to helium $1s^12p^1$. The energy of $|\varphi_{1}\rangle$, $\omega_{1}=15.7$ eV, is chosen such that the resonance-transition energy between $|\varphi_{1}\rangle$ and $|\varphi_{2}\rangle$ is larger than the dressing field frequency $\omega_{L}=1.38$ eV in the IR region, for fulfilling the adiabatic conditions. The transition matrix element from the ground state to $|\varphi_{2}\rangle$ is $d_{g2}=0.5$, while the transition matrix element from $|\varphi_{1}\rangle$ to $|\varphi_{2}\rangle$ is $d_{12}=2.7$ being much larger than $d_{g2}$. In order to verify the validity of the adiabatic approximation, the time-dependent Schr\"{o}dinger equation for the three-level system is numerically solved as well, including the effects of the pump-probe pulses field on evolution processes of the system. The pulse duration of the IR-dressing field is $\tau=13.34$ fs, with a strength of $E_{0}=0.02\, {\rm a.u.}$($\approx 1.4\times 10^{13}\, {\rm W/cm}^{2}$), while the duration of the attosecond XUV pulse with an intensity of $1\times 10^{10}\, {\rm W/cm}^{2}$, centered in at $21.4\, {\rm eV}$, is $240\, {\rm as}$. The instant-changed $\Delta C$ that results from the attosecond pulse is set to be $0.05$.

The delay-dependent response function $S(\omega,t_{d})$ calculated from numerical solution is shown in Fig.~\ref{f2}(a). In the figure positive delays correspond to the attosecond pulse arriving after the center of the IR-dressing field. We observe a strong absorption line at $\omega_{2}$ in the spectra, corresponding to the resonance transition from the ground state to $|\varphi_{2}\rangle$ induced by the attosecond pulse. The spectra also exhibits the interference features, which are distributed symmetrically around $\omega_{1}\pm\omega_{L}$ and $\omega_{2}\pm 2\omega_{L}$, respectively. The sideband signals  in spectra around $\omega_{1}$ and $\omega_{2}$ are related to the quasi-harmonics from the dipole response of the laser-dressed states. The spectrum of the quasi-harmonics is introduced in the Ref.~\cite{23,34}, here it can be attributed to multi-photon-dressed processes of the IR-dressing field with the three-level system. The values of the sideband signals depend on the coherent interaction between the quasi-harmonics and the attosecond pulse. For analysis the quasi-harmonics, we calculate the laser-dressed dipole response by the adiabatic model, as shown in Fig.~\ref{f2}(b), where the adiabatic model are used to calculate the dipole response for the same parameters as in Fig.~\ref{f2}(a). We find that the adiabatic model yields a good agreement with the numerical solution of the three-level TDSE.

Because of the different energies of the quasi-harmonics from different dressed states, the spectra is classified into two parts by energies: $\omega_{2}\pm2\omega_{L}$ and $\omega_{1}\pm\omega_{L}$, as shown in Fig.~\ref{f2}(b1,b2). The dipole responses of the two parts, corresponding to the transitions from the dressed states $|\varphi_{\pm}\rangle$ to the ground state with transition energies $\omega_{1}+\omega_{\pm}$, are calculated by the two terms of $C_{2}(t)$ respectively in Eq.~(\ref{e3}): $C_{\pm}a_{\pm}e^{-i[(t-t_{d})\cdot\omega_{1}+\int \omega_{\pm}dt]}$, for $d(t)\propto Re[e^{-i(t-t_{d})\cdot\omega_{1}}C_{2}(t)]$ (see Appendix~\ref{Ap1}). We find the classifying of the spectra is helpful to analyse the interference features in the spectra. In next Sec.~\ref{S31} and~\ref{S32}, we will discuss the two main delay-dependent features in Fig.~\ref{f2}: The interference fringes from two-photon-dressed processes as a function of delay and photon-energy, and the modulation signals of the laser-induced sidebands of the dark state $|\varphi_{1}\rangle$ at $\omega_{1}\pm\omega_{L}$ in the spectra when the pump-prop fields overlap($t_{on}\leq t_{d}\leq t_{off}$).

\subsection{Delay-dependent phase of interference fringes from two-photon-dressed processes}\label{S31}
The interference fringes around $\omega_{2}\pm2\omega_{L}$ in delay-dependent absorption spectra result from the interference between direct and indirect pathways, which has been discussed previously~\cite{6,11}, and we will further study the phase of the interference fringes in more details in this section. According to the response function $S(\omega,t_{d})$, the absorption of the XUV field pulse could be regarded as a coherent interaction processes depending on the phase difference between the XUV field pulse and the laser-dressed dipole response of the atom. In the following, we extract the quasi-harmonics from the laser-dressed dipole response, which can contribute to the coherent interaction processes around the energy $\omega_{2}\pm2\omega_{L}$. Therefore, the delay-dependent phase of the interference fringes is discussed by the coherence-phase difference between the attosecond XUV pulse and the quasi-harmonics.

\begin{figure}
\centering
\begin{minipage}[htb]{0.5\textwidth}
\centering
\centerline{\includegraphics[width=3.0in]{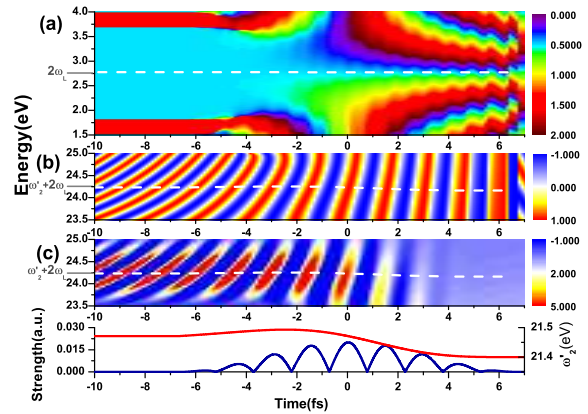}}
\end{minipage}
\caption{(color online) The interference fringes at $\omega'_{2}+2\omega_{L}$. (a) The phase offset $\phi$ as a function of photon energy and delay. The pixel value is measured in $\pi$. (b) The interference fringes calculated by Eq.(\ref{e011}), where $\phi$ is given in (a) and $\omega'_{2}$ is shown as the red line. (c) The interference fringes being same in Fig.~\ref{f2}(a).
\label{f3}}
\end{figure}

The coherence-phase difference between the attosecond XUV pulse and the quasi-harmonics from two-photon-dressed processes is required for describing the response function $S(\omega,t_{d})$ around $\omega_{2}+2\omega_{L}$ in the ATA spectra. The sketch map of generation processes of the coherence-phase difference is shown in Fig.~\ref{f2}(c). The spectral phase of the attosecond XUV pulse can be obtained by Fourier transform. In order to get the phase of the dipole response, we assume there is a $\frac{\pi}{2}$ phase difference between the attosecond pulse and the dipole response at the resonance-transition energy $\omega_{2}$, that is reasonable for the resonance absorption by the system. But there need a modification on the resonance-transition energy $\omega_{2}$ for the as-stark shift by the IR-dressing field~\cite{14,41}. We consider the effect of the time-averaged shift energy $\delta\omega$, over the part duration of the IR-dressing field pulse arriving after $t_{d}$. Then the resonance-transition energy is approximated by:
\begin{equation}\label{e013}
\omega'_{2}\approx\omega_{2}+\delta\omega(t_{d}),
\end{equation}
where $t_{on}<t_{d}\leq t_{off}$, and the shift energy $\delta\omega(t_{d})$ is given in Eq.~(\ref{a1}) according to the adiabatic model (see Appendix~\ref{Ap2}). The modified energy $\omega'_{2}$ as a function of delay is shown in the red line of Fig.~\ref{f3}. The excited dipole response propagates in delay period, which causes a phase difference $(\omega-\omega'_{2})t_{d}$ for the different frequencies. After the delay period, the excited dipole is dressed by the dressing field, and generates quasi-harmonics with energy $\omega'_{2}\pm 2\omega_{L}$. The quasi-harmonics, corresponding to the effects of the two-photon of the IR-dressing field, have a phase difference $\phi-\frac{\pi}{2}$ with the dipole response at $\omega'_{2}$, where the phase offset $\phi$ is given by~(see details in Appendix~\ref{Ap2}):
\begin{equation}\label{e012}
 \phi(\omega,t_{d})=arg[\int^{+\infty}_{-\infty}\theta(t)e^{-i\omega t}dt].
\end{equation}
The phase offset $\phi$ arising from the coupling processes between the IR-dressing field and the system is shown in Fig.~\ref{f3}(a).

The coherence-phase difference is $(\omega-\omega'_{2})t_{d}+\phi$, and then the response function $S(\omega,t_{d})$ can be given as a function of the coherence-phase difference approximately:
\begin{equation}\label{e011}
  S(\omega,t_{d})\propto \sin{[(\omega-\omega'_{2})t_{d}+\phi(\omega-\omega'_{2},t_{d})]}.
\end{equation}
With the function, we can describe delay-depended interference fringes, as shown in Fig.~\ref{f3}.
The phase offset $\phi(\omega,t_{d})$ and $\omega'_{2}(t_{d})$ are unchanged when attosecond pulse before the IR-dressing field $t_{d}<t_{on}$, and the shapes of the interference fringes follow the hyperbolic lines. When $t_{on}<t_{d}\leq t_{off}$ the pump-probe field overlaps, the fringe shapes are changed dramatically, and the slope of the fringes will saturate when delay time gets close to the end of the dressing field. The changes of the fringe shapes, depended on the coherence-phase difference as a function of delay and energy, are obviously indicated by the varies of the phase offset $\phi(\omega,t_{d})$, especially before the peak of the IR-dressing field. We also find $\phi(2\omega_{L})=\frac{\pi}{2}$ is independent on the delay time, which represents the two-photon-dressed processes of the dressing field varies little with respect to delay. It indicates the strong absorption of the attosecond pulse at $\omega'_{2}+2\omega_{L}$ occurs at the peak times of the dressing field. That can be utilized to calibrate the position of the interference fringes in the delay-dependent spectra.

We remake here the adiabatic model could capture much of the three-level system dynamics with the resonance-transition energy $\Delta\omega_{12}$ larger than the photon energy $\omega_{L}$ of the dressing field, but the dynamics of the adiabatic model cannot include transition processes between the dressed states. Actually, the transition processes cannot be ignored when $|\Delta\omega_{12}|\leq\omega_{L}$, which may affect the laser-dressed dipole response phase and the interference fringes in ATA spectra as well. Although the adiabatic model is not good enough, it could give a quantitative description of the interference features from the laser-dressed system in ATA.

\subsection{Modulation signals on sidebands from one-photon-dressed processes}\label{S32}

Compared to two-photon-dressed processes in Sec.~\ref{S31}, the modulations of the sidebands with one-photon coupled processes demonstrate different interference-fringe feature. In this section, we focus on the modulation of the spectra of the laser-induced sidebands with energy $\omega_{1}\pm \omega_{L}$ in the spectra, corresponding to one-photon-dressed processes, as shown in Fig.~\ref{f2}(b2). Actually, the sidebands are strongly affected by the pump probability of the dressed state $|\varphi_{-}\rangle$ excited by the attosecond pulse.

\begin{figure}
\centering
\begin{minipage}[h]{0.5\textwidth}
\centering
\centerline{\includegraphics[width=3in]{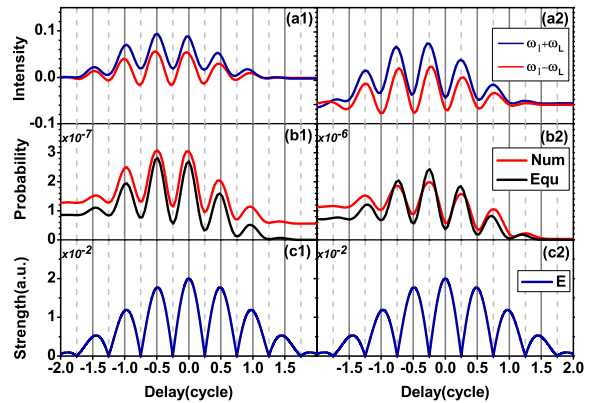}}
\end{minipage}
\caption{(color online) Comparison of the modulation phase of the sidebands of state $|\varphi_{1}\rangle$ with $|\Delta \omega_{12}| \gg \omega_{L}$(left,a1-c1) and $|\Delta\omega_{12}| \ll \omega_{L}$(right,a2-c2). The $\Delta \omega_{12}$ is changed by setting the energy value of $|\varphi_{1}\rangle$ with $\omega_{1}=15.7eV$ and $21.0eV$. The delay time is given in cycle time of the dressing field for clarity. (a) Area of the sideband signal (integrated over the range of 0.5eV with the center energy at the sideband peak intensity in Fig.~\ref{f2}(a)) as a function of the delay time. (b) Average occupancy $\overline{|C_{1}|^2}$ given in Eq.~(\ref{e112}) as a function of the delay time. Red Lines with label $Num$ are numerically solved by TDSE, and black lines with label $Equ$ are by Eq.~(\ref{e004}). (c) Absolution of the dressing field strength as a function of the delay time for reference.
\label{f4}}
\end{figure}

In the adiabatic model, the resonance absorption of the attosecond pulse at $\omega_{1}\pm \omega_{L}$ can be related to the occupation probability of the $|\varphi_{-}\rangle$. Considering the relations between the dressed states and the field-free states from Eq.~(\ref{e2}), the changes of occupancy of $|\varphi_{1}\rangle$ reflect the occupation probabilities of $|\varphi_{-}\rangle$. Then the intensity of the sidebands can be indicated by the average of $|\varphi_{1}\rangle$ occupancy over the dressing field pulse duration:
\begin{equation}\label{e112}
  \overline{|C_{1}|^2}=\frac{1}{t_{off}-t_{on}}\int^{t_{off}}_{t_{on}}|C_{1}(t)|^{2}dt,
\end{equation}
where $\overline{|C_{1}|^2}$ is the average of occupation probability of $|\varphi_{1}\rangle$ over the dressing field pulse duration. The modulation signals of the sidebands of $|\varphi_{1}\rangle$ can be indicated by the parameter $\overline{|C_{1}|^2}$, as shown in Fig.~\ref{f4}, where the lines of the dressing field strength, average occupancy $\overline{|C_{1}|^2}$, and the sidebands signals demonstrate the identical modulations. Here, in adiabatic condition $|\Delta\omega_{12}|\gg\omega_{L}$, the transition to $|\varphi_{1}\rangle$ depends on the coupling between the field-free states by the dressing field. The strength of the dressing field is strong at $t_{d}$ and then strongly couples the field-free states, that makes the transition to $|\varphi_{1}\rangle$ with "XUV+IR" processes occurring more easily. Such that the population of $|\varphi_{1}\rangle$ varies, periodically with respect to the delay time, as the changes of strengths of the dressing field at the pump times of the attosecond pulse.

We also calculate the case with $|\Delta\omega_{12}|\ll\omega_{L}$. With the condition, the strongly Rabi flopping between the bound states induced by the dressing field will join in the dipole oscillation of the system. It causes the vibration of the populations of the dressed states and results in the non-adiabatic processes. The calculations of the case are shown in Fig.~\ref{f4}. And we find that there is a $\pi$ phase difference with the lines from the adiabatic case from Fig.~\ref{f4}.

The $\pi$ phase shift of $\overline{|C_{1}|^2}$ in non-adiabatic processes is related to the effects of the dressing field after $t_{d}$, which bring on the strongly Rabi flopping. To make the points more clear, we calculate the average occupancy $\overline{|C_{1}|^2}$, by only considering the transition from the $|\varphi_{2}\rangle$ to $|\varphi_{1}\rangle$ with the effects of the dressing field. Here we assume that the dressing field causes little change of the probability amplitude of $|\varphi_{2}\rangle$ with the off-resonant condition, such that $C_{2}(t)\approx \Delta C \exp{[-i(t-t_{d})\cdot\Delta\omega_{12}]}$. Then the average occupancy $\overline{|C_{1}|^2}$ is (see Appendix~\ref{Ap3}):
\begin{equation}\label{e004}
  \begin{array}{l}
    \overline{|C_{1}|^2}\approx \frac{(t_{off}-t_{d})\eta^{2}\varepsilon(t_{d})^{2}}{t_{off}-t_{on}}\big[\omega_{L}^{2}\sin^{2}{(\omega_{L}t_{d})}+ \\
    \Delta\omega^{2}_{12}\cos^{2}{(\omega_{L}t_{d})}\big]+\eta^{2}\int^{t_{off}}_{t_{d}}\varepsilon(t)^{2}\frac{\omega_{L}^{2}+\Delta\omega^{2}_{12}}{2}dt,
  \end{array}
\end{equation}
where $t_{on}\leq t_{d}\leq t_{off}$. The results calculated by Eq.~(\ref{e004}) fit well with the numerical results, in the off-resonant conditions, as shown in Fig.~\ref{f4}. The $\pi$ phase difference validates the changes of the effects of the dressing field in the two off-resonant conditions. Note that the signals of the sidebands of $|\varphi_{1}\rangle$ are sensitive to $\overline{|C_{1}|^2}$ in Fig.~\ref{f4}, even though the modulation changed with the different conditions. This shows the modulation phases of the sideband signals with one-photon-dressed processes are also affected by the following dressed processes after the system excited by the attosecond pulse.

\section{Summary}\label{S4}  

In conclusion, we have investigated the interference features of the ATA spectra by the three-level system in the present of both the IR-dressing field and a delayed attosecond pulse. We have analysed the interference fringes with the adiabatic model and compared it with the numerical solution of the three-level TDSE. The model revealed the qualitative relationship between of the interference fringes and the dressing field. Our results showed the coherent interaction processes between the attosecond pulse and the quasi-harmonics from the two-photon-dressed processes could result in the interference fringes emerged in the ATA spectra. The interference fringes are determined by the coherence phase difference, which is dependent on the pump-probe delay time and the energy-shift processes driven by the dressing field. We also showed the signals of the laser-induced sidebands around the dark level in the ATA spectra is associated with the average occupancies of the dark state over the duration of the dressing field, that are modulated periodically with respect to the delay time. Moreover if the photon energy of the dressing field strides across the resonance-transition energy between the bound states, the laser-induced sideband signals as a function of delay will have a $\pi$ phase jump. That showed the modulations of the sideband signals in the ATA spectra are also affected by the following processes dressed by the IR laser field after the attosecond pulse. We have studied the roles of quasi-harmonics extracted from the IR-dressed dipole response played in the ATA processes, expect that it conduces to understanding of the "XUV+2IR" and "XUV+IR" transition processes.

\section{ACKNOWLEDGMENTS}
This work is supported by the National Basic Research Program of China (973 Program)
under Grant No. 2013CB922203, the NSF of China (Grants No. 11374366 and No. 11274383), and the National High-Tech ICF Committee of China

\appendix
\section{solutions of adiabatic model}\label{Ap1}

With the transform relation between the dressed states and the dark states $|\varphi_{1}\rangle$ from Eq.~(\ref{e2}), Eq.~(\ref{eq:e1}) can be expressed as in dressed states representation:
\begin{equation}\label{eq:e4}
  \left\{
  \begin{array}{c}
    i\frac{\displaystyle dC_{-}}{\displaystyle dt}=\omega_{-}C_{-}-i\beta_{+}C_{+} \\
     \\
    i\frac{\displaystyle dC_{+}}{\displaystyle dt}=\omega_{+}C_{+}-i\beta_{-}C_{-}
  \end{array}
  \right.
\end{equation}
When it satisfies the adiabatic condition: $|\Delta\omega_{12}|\gg\omega_{L}$ and the intensity of the dressing field is weak($\leq 10^{14}$ W/cm$^{2}$), such that $|\beta_{\pm}|=|\frac{\gamma\prime\Delta \omega_{12}}{4\gamma^{2}+\Delta\omega^{2}_{12}}|\ll 1$, then the system is evolving adiabatically~\cite{37,38}. In the adiabatic approximation, we take $|\beta_{\pm}|=0$.

Assuming at $t_d$, the attosecond pulse induces an instant change of the probability amplitude $\Delta C$ of $|\varphi_{2}\rangle$, then the probability amplitudes of the two dressed states are given by $C_{\pm}(t_{d})=\Delta Ca_{\pm}(t_{d})$. In the subsequent evolution, this equation has the following solutions:
\begin{eqnarray}\label{a3}
\left\{\begin{array}{c}
         C_{2}(t)=C_{-}(t_{d})a_{-}(t)\exp(-i\int^{t}_{t_{d}}\omega_{-}dt)+\\C_{+}(t_{d})a_{+}(t)\exp(-i\int^{t}_{t_{d}}\omega_{+}dt), \\
         C_{1}(t)=C_{-}(t_{d})b_{-}(t)\exp(-i\int^{t}_{t_{d}}\omega_{-}dt)+\\C_{+}(t_{d})b_{+}(t)\exp(-i\int^{t}_{t_{d}}\omega_{+}dt).
       \end{array}
\right.
\end{eqnarray}
In particular, this paper focuses on dipole response from $|\varphi_{2}\rangle$ to the ground state with respect to the pump-probe delay. The dipole response in time domain is given by: $d(t)\approx 2Re[e^{-i(t-t_{d})\cdot\omega_{1}}C_{2}(t)d_{g2}]$, where it is assumed the probability amplitude of the ground state changes little, and the $C_{2}(t)$ is no transition from the ground state before $t_{d}$ and then varies as described by Eq.~(\ref{a3}) after $t_{d}$.

\section{phase of quasi-harmonics with two-photon-dressed processes}\label{Ap2}
With the adiabatic model, $C_{+}a_{+}e^{-i[(t-t_{d})\cdot\omega_{1}+\int \omega_{+}dt]}$ in Eq.~(\ref{e3}) contributes to the dipole response with two-photon-dressed processes. Under the adiabatic condition, $\omega_{+}\approx\Delta\omega_{12}$, allow to make a approximation of this term:

\begin{equation}\label{e01}
\begin{array}{l}
C_{+}(t_{d})a_{+}(t)e^{-i[(t-t_{d})\cdot\omega_{1}+\int^{t}_{t_{d}} \omega_{+}dt']}\\\approx C_{+}(t_{d})e^{-i[(t-t_{d})\cdot\omega_{2}+\int^{t}_{t_{d}}\frac{\gamma^2(t')}{\Delta\omega_{12}}dt']}\\
\approx C_{+}(t_{d})e^{-i[(t-t_{d})\cdot\omega_{2}+\delta\theta(t)]}e^{-i\theta(t)},\\
\end{array}
\end{equation}

where the $\delta\omega$ and $\theta(t)$ is defined as:
\begin{equation}\label{e03}
\begin{array}{l}
 \delta\theta(t)=\int^{t}_{t_{d}}\frac{[E_{0}d_{12}\varepsilon(t')]^2}{2\Delta\omega_{12}}dt', \\
 \theta(t)=\int^{t}_{t_{d}}\frac{\gamma^2(t')}{\Delta\omega_{12}}dt'-\delta\theta(t).\nonumber
\end{array}
\end{equation}
Average of the shift energy over $[t_{d},t_{off}]$ is given in:
\begin{equation}\label{a1}
\delta\omega(t_{d})=\frac{\delta\theta(t_{off})}{t_{off}-t_{d}},
\end{equation}

For extracting the phase of the quasi-harmonics with two-photon-dressed processes from the dipole response, the factor $e^{-i\theta(t)}$ is expanded in a Taylor series with the condition $\left|\theta(t)\right|< 1$:
\begin{equation}\label{e02}
  e^{-i\theta(t)}\approx 1-i\theta(t),
\end{equation}
where only keep the first two terms, and the high-order terms include high-order effects of the IR-dressing field are neglected. The two terms of Eq.~(\ref{e02}) relate to the dipole response from $|\varphi_{2}\rangle$ to the ground, and the quasi-harmonics with two-photon-dressed processes of $|\varphi_{2}\rangle$, respectively. The quasi-harmonics gets a phase difference ($\phi-\frac{\pi}{2}$) with the dipole response of the level $|\varphi_{2}\rangle$, where $\frac{\pi}{2}$ comes from the factor i in Eq.~(\ref{e02}). According to the Eq.~(\ref{e02}), the phase offset $\phi$ is given by:
\begin{equation}\label{a012}
 \phi(\omega,t_{d})=arg[\int^{+\infty}_{-\infty}\theta(t)e^{-i\omega t}dt],
\end{equation}

\section{occupancy of $|\varphi_{1}\rangle$ with off-resonant condition}\label{Ap3}
Considering the transition from the $|\varphi_{2}\rangle$ to $|\varphi_{1}\rangle$, where assume that the dressing field causes little change of the probability amplitude of $|\varphi_{2}\rangle$ with the off-resonant condition, such that $C_{2}(t)\approx \Delta C \exp{[-i(t-t_{d})\cdot\Delta\omega_{12}]}$, the probability amplitude of the field-free state $|\varphi_{1}\rangle$ is:
\begin{equation}\label{e002}
 C_{1}(t)=-i\Delta C\int_{t_{d}}^{t}E(t')d_{12}e^{-i(t'-t_{d})\cdot\Delta\omega_{12}}dt',
\end{equation}
Then according to Eq.~(\ref{e112}), the integral average $\overline{|C_{1}|^2}$ is:
\begin{equation}\label{a004}
  \begin{array}{l}
    \overline{|C_{1}|^2}\approx \frac{(t_{off}-t_{d})\eta^{2}\varepsilon(t_{d})^{2}}{t_{off}-t_{on}}\big[\omega_{L}^{2}\sin^{2}{(\omega_{L}t_{d})}+ \\
    \Delta\omega^{2}_{12}\cos^{2}{(\omega_{L}t_{d})}\big]+\eta^{2}\int^{t_{off}}_{t_{d}}\varepsilon(t)^{2}\frac{\omega_{L}^{2}+\Delta\omega^{2}_{12}}{2}dt,
  \end{array}
\end{equation}
where $\eta=\frac{E_{0}d_{12}\Delta C}{\Delta\omega^2_{12}-\omega_{L}^2}$, $t_{on}\leq t_{d}\leq t_{off}$.


\newpage

\end{document}